      [1999/12/01 v1.4c Il Nuovo Cimento]
\documentclass[varenna]{cimento}
\usepackage{graphicx,amssymb,amsmath,citesort}

\title{Localisation versus self-trapping: Polaron formation in the Anderson-Holstein model}

\author{Holger~Fehske, Franz~X.~Bronold, \atque Andreas~Alvermann}

\institute{Institut f\"ur Physik, 
Ernst-Moritz-Arndt-Universit\"at Greifswald,
17487 Greifswald, Germany}
\shortauthor{H. Fehske, F.-X.~Bronold, \atque A.~Alvermann }

\begin{document}

\maketitle

\section{Problem}

The polaron concept, first introduced by Landau~\cite{La33}, is one of the main 
pillars on which the theoretical analysis of materials with strong electron-phonon
(EP) coupling rests. In these compounds, the coupling between the electron and the 
lattice leads to a lattice deformation whose potential tends to bind the 
electron to the deformed region of the crystal. This  process, which has been called 
self-trapping because the potential depends on the state of 
the electron, does not destroy translational invariance, even if the
lattice deformation is confined to a single lattice site (small
polaron)~\cite{FE,FAHW_Varenna}. Quantum mechanical tunnelling 
between different lattice sites restores this symmetry and ensures
that a self-trapped electron forms an itinerant polaronic quasi-particle-particle.

In contrast to a weakly phonon-dressed electron, a polaron is strongly
 mass-enhanced, because of the lattice distortion it has to coherently
 carry along.
  A polaron is therefore rather sluggish and 
accordingly susceptible to random changes in the local environment.  
Crystallographic and chemical defects may easily turn an itinerant polaron 
into a localised particle which is confined to a finite region of the crystal.
This confinement can arise as the polaron binds to an impurity
which is energetically separated from the band of itinerant states.
More interestingly the scattering on defects can lead to destructive
interference which transmutes an itinerant state to a localised state.
This transmutation is called Anderson localisation~\cite{An58,KK93}. It 
is as central to the analysis of disordered materials as the polaron concept 
is for materials with strong EP coupling. 

An electron in a localised state is of course 
strongly coupled to local lattice distortions. Anderson localisation and self-trapping 
are thus intricately connected. Yet, a combined analysis of the two is, with a 
few exceptions~\cite{An72,GJ80,MT83,CES83}, conspicuously missing in the 
polaron literature. This is a serious shortcoming because a priori it is not clear whether 
Anderson localisation and self-trapping always work in the same direction, as suggested 
by the simple discussion given above. 

The purpose of this contribution is to describe a theoretical approach capable of 
accounting for both Anderson localisation and self-trapping,
 and to apply it to the analysis of the issue just mentioned. 
We present  
precise criteria to distinguish itinerant from localised polarons.
With these criteria at hand we 
study the localisation behaviour of a single polaron 
in dependence on the EP coupling.

\begin{figure}[t]
\begin{minipage}{0.5\linewidth}
\includegraphics[width=\linewidth,clip]{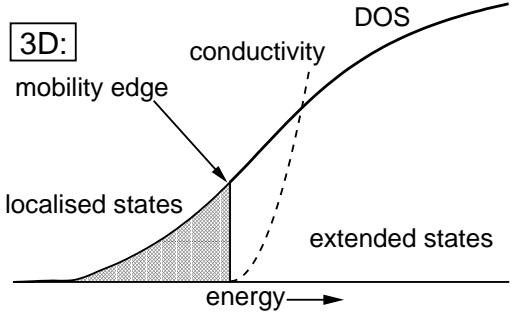}
\end{minipage} \hfill
\begin{minipage}{0.45\linewidth} \vspace*{.5cm}  
\caption{\small The figure displays the density of states (DOS)
for the single-electron Anderson model. States deep inside the band are 
extended, whereas states below (above) the lower (upper) mobility edge  
are localised. The dc-conductivity vanishes at the
mobility edge. Note that the DOS is finite in the region of
localised states.} 
\label{Illus}
\end{minipage}
\end{figure}

\section{Modelling}

To discuss the interplay of EP coupling and 
randomness, let us consider, as a generic model, the single-particle Anderson-Holstein 
Hamiltonian
\begin{equation}
      H = \sum\limits_{i} \epsilon_i n_{i}
      - t
      \sum\limits_{\langle i, j\rangle}  c_{i}^\dagger c_{j}^{} 
      - \sqrt{\varepsilon_p\omega_0} 
      \sum\limits_{i} (b_i^\dagger + b_i^{} ) n_{i}
      + \omega_0 \sum\limits_i b_i^\dagger b_i^{}~,   
   \end{equation} 
where $t$ is the electron transfer amplitude between neighbouring
      sites on a three-dimensional (3D) lattice, $\omega_0$
is the bare phonon frequency ($\hbar=1$), $\varepsilon_p$ is the polaron shift, and
$n_i=c_i^\dagger c_i$ is the local charge density. 
The on-site energies $\{\epsilon_i\}$ are independently identically
      distributed  
random variables with common distribution 
$P( \epsilon_i)  = \frac{1}{\gamma}
              \theta\left(\frac{\gamma}{2}-|\epsilon_i|\right)$.

Without EP coupling ($\varepsilon_p=0$), i.e. for
a bare electron in a disordered 3D crystal,
Anderson has shown that all states are localised 
provided the width $\gamma$ of the distribution 
of the on-site energies $\epsilon_i$ exceeds a critical value
$\gamma_c$~\cite{An58}. 
When $\gamma<\gamma_c$, electron states towards 
the band edges would be localised in this model whereas states towards
the band centre would be extended (and thus itinerant). Mobility edges at 
$\pm \omega_\mathrm{mob}(\gamma)$ separate the
two classes of electron states~\cite{Mo66} (see Fig.~\ref{Illus}). 
We will be only concerned with this 3D setting:
In 1D (and arguably 2D) 
tight-binding models with random on-site energies all states are
localised no matter how small $\gamma$.

\section{Stochastic Green's function approach}

The most natural quantity to characterise the localisation of 
a bare electron is the localisation length $l$ 
defined by $|\psi({\bf r})|\propto \exp{(-r/l)}$, 
where $\psi({\bf r})$ is the electron wave-function at energy $\omega$. 
If $l$ diverges (is finite), the state is extended (localised). In most
cases of interest, however, the wave-function cannot be straightforwardly 
determined.
Alternative quantities suitable for a theoretical analysis 
are the (disorder averaged) $T=0$ dc-conductivity which is finite
only for extended states, or the $T=0$ return-probability
which is finite only for localised states. 
Both quantities are four-point correlation functions and thus also hard to obtain. 
Two-point functions, for instance the local density of states (LDOS),
\begin{eqnarray}
\rho_i(\omega)=\sum_n|\psi_n({\bf r}_i)|^2\delta(\omega-E_n)
=-\frac{1}{\pi}{\rm Im}\,G_{ii}(\omega)
\end{eqnarray}
($G_{ii}(\omega)$ is the local Green's function), 
are much easier to calculate but then disorder averaging is inappropriate:
The (arithmetically averaged) DOS
\begin{eqnarray}
\rho_\mathrm{av}(\omega) = \langle \rho_i(\omega) \rangle = \int \rho_i P[\rho_i(\omega)] \,
d\rho_i
\end{eqnarray}
gives the number of states at $\omega$ independent of whether
they are extended or localised.

It was Anderson who pointed out that in a disordered system one must
study the {\it distribution} of such two-point functions~\cite{An58}.
From the distribution $P[\rho_i(\omega)]$ of the LDOS
one can decide whether the state at energy $\omega$ is localised or not. 
For the ordered system $\rho_i(\omega)$ does not depend on $i$ due to
translational symmetry, and $P[\rho_i(\omega)]=\delta[\rho_i-\rho(\omega)]$.
When, in an extended state,  the electron wave-function has more or
less equal weight on every lattice site, $P[\rho_i(\omega)]$ is rather
symmetric and usually Gaussian.
With increasing disorder, the weight of the wave-function becomes more and 
more restricted to a few lattice sites leading to a broad and
asymmetric distribution (see left lower panel in Fig.~\ref{MEelectron}
below).
Above the localisation transition, the wave-function has appreciable
weight only in a finite region.
Thus, the  distribution $P[\rho_i(\omega)]$ is singular,
i.e. $P[\rho_i(\omega)]  = \delta[\rho_i]$,
signalling a localised state at energy $\omega$.  

To analyse the effects of disorder on polaron formation on the level of 
two-point functions we applied the recently developed statistical dynamical 
mean field theory (statDMFT)~\cite{DK97} to the Anderson-Holstein model. The statDMFT 
is a stochastic Green's function approach, based on the self-consistent construction of 
random samples (and thus distributions) for local Green's functions. 
Conceptually, it is an extension of the self-consistent theory of 
localisation of bare electrons~\cite{AAT73} to models with interactions
and has been successfully applied to the disordered Hubbard~\cite{DK97,DK98},
Anderson lattice~\cite{AMD03}, and Holstein models~\cite{BF02,BAF04}. 

\begin{figure}[t]
\hspace{0.05\linewidth}\begin{minipage}{0.25\linewidth}
\includegraphics[width=\linewidth]{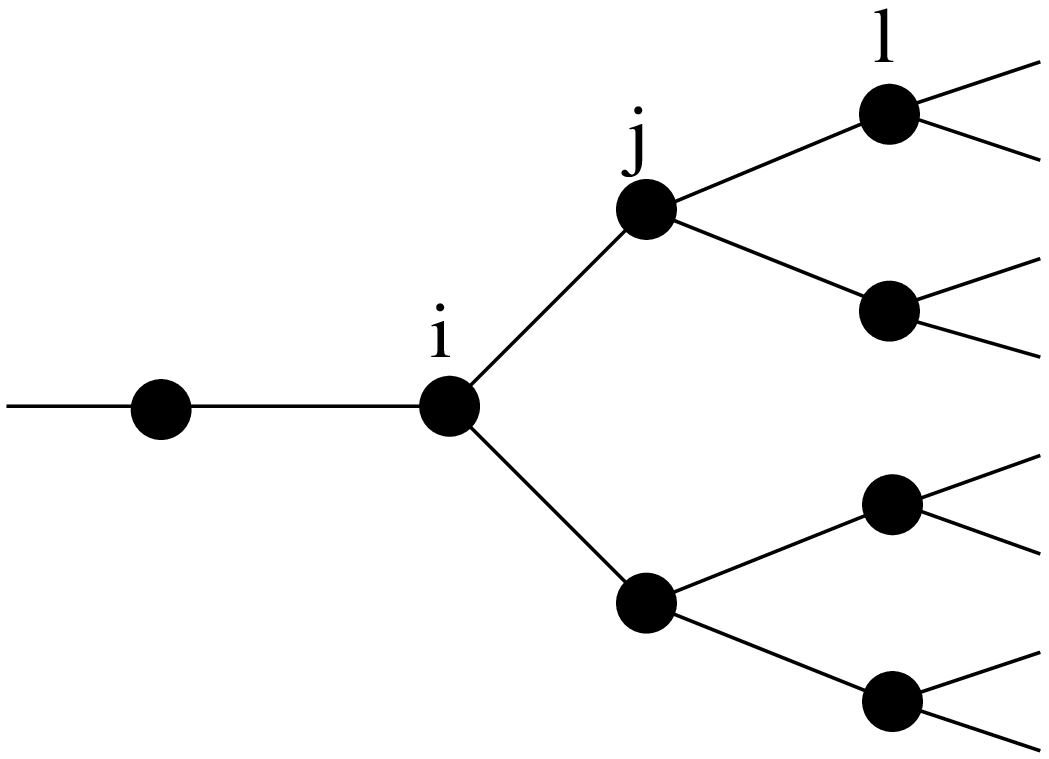}
\end{minipage}\hspace{0.05\linewidth}
\begin{minipage}{0.55\linewidth}
\includegraphics[width=\linewidth]{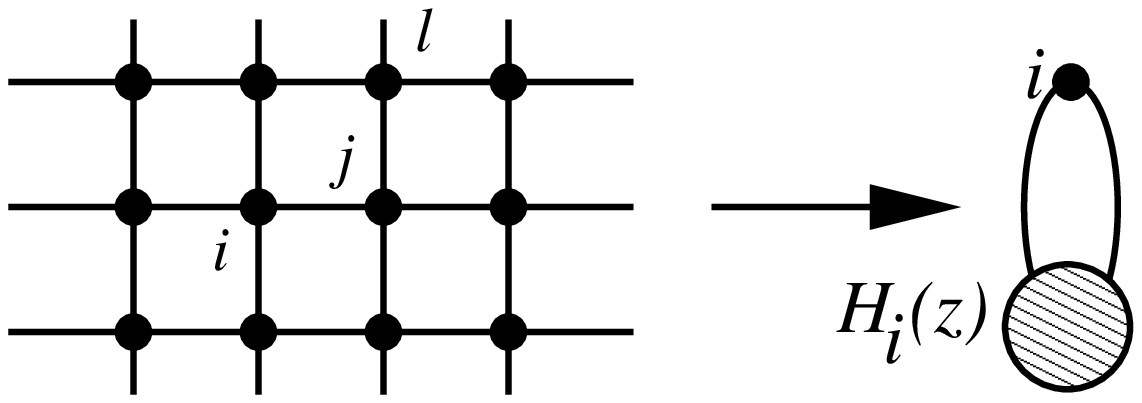}
\end{minipage}
\caption{\small Philosophy of the statDMFT: The lattice
problem is mapped on to an ensemble of impurity problems, each consisting of
a single site $i$ self-consistently embedded in an effective medium described
by the hybridisation function $H_i(z)$; all lattice information
is contained in this function. For a Bravais lattice the mapping leads 
to rather complicated equations. The statDMFT construction is therefore
done for a Bethe lattice, shown on the very far left,
where the fact that site $i$ cannot be reached from $l$ once $j$ is
removed leads to dramatic simplifications.}
\label{statDMFT}
\end{figure}

The statDMFT maps the original lattice on to an ensemble 
of impurity sites each of which is self-consistently embedded in an
effective medium comprising the environment of 
this site (cf. Fig.~\ref{statDMFT}). The main 
approximation is to treat interaction processes 
as in conventional dynamical mean field theory (DMFT).
Within this approximation EP coupling gives rise to a local  
self-energy $\Sigma_i(z) $ only. For a single electron in the Anderson-Holstein
model $\Sigma_i(z)$ becomes a continued fraction~\cite{Su74CDF97}
\begin{equation}\label{syk} 
\Sigma_i(z) = \cfrac{1 \varepsilon_p \omega_0}
{z-1 \omega_0-\epsilon_i-H_{i}(z-1 \omega_0) - 
\cfrac{2 \varepsilon_p \omega_0}{z-2 \omega_0-\epsilon_i-H_{i}(z-2 \omega_0) - 
\cfrac{3 \varepsilon_p \omega_0}{\cdots}}
}
\end{equation}
with $z=\omega+i\eta$ and the hybridisation function $H_i(z)=t^2
 \sum_{j}G_{jj}(z)$.
The index $j$ runs over the $K$  sites neighbouring to $i$,
when $K$ is the connectivity of the lattice. 
The $K$ Green's functions $G_{jj}(z)$  in this sum have to be evaluated
for a lattice with site $i$ removed.

With this self-energy the local Green's 
function $G_{ii}(z)$ of the single-electron Anderson-Holstein model at $T=0$ is given by
\begin{equation}
G_{ii}(z)=
\frac{1}{\displaystyle z-\epsilon_i-H_i(z)-
\Sigma_i(z)} \label{Gii}
\end{equation}
with a local contribution $\epsilon_i-\Sigma_i(\omega)$, and $H_i(z)$
accounting for the hopping to neighbouring sites.
Apart from the DMFT treatment of interaction these equations
are exact on a Bethe lattice where the removal
of sites preserves the lattice structure.
The details of the derivation are given in Ref.~\cite{BAF04}.

The ensemble of impurity sites is then analysed in terms of
probability distributions.
Due to the randomness of the on-site energies, 
$G_{ii}(z)$ is a random variable whose 
distribution function has to be numerically determined from 
Eq.~(\ref{Gii}). Towards that end, we proceed as follows: 
First, we reinterpret the site indices as labels
enumerating the $N$ elements of a random sample; each $G_{ii}(z)$,
$i=1,\cdots,N$, can be understood as a particular realisation
of a random variable.
We then self-consistently construct 
the random sample $\{G_{ii}(\omega)\}$, 
starting from an initial random sample, which we successively update
via a Monte Carlo sampling algorithm, drawing the random variables on the right habd side of
Eq.~(\ref{Gii}) from the corresponding random samples created the iteration step before.
Because in Eq.~(\ref{syk}) $G_{jj}(\omega-p\omega_0)$ appears
we have to simultaneously construct a random sample for all 
$G_{ii}(\omega-p\omega_0)$, with
$p=0,\cdots,M$, where $M$ is the truncation depth of the continued fraction.

For reasonable numerical accuracy,
the sample size $N$, which should not be confused with the actual size of the Bethe
lattice but instead gives the precision with which 
we construct the random sample, has to be sufficiently large. Typical sample sizes  are
$N\approx 50~000$. The depth $M$ of the continued fraction, which describes
the maximum number of virtual phonons in the lattice, has to be large enough in
order to capture polaron formation. As a rough estimate we use
$M\approx 5g^2$, where $g^2=\varepsilon_p/\omega_0$ is approximately the
average number of virtual phonons comprising the phonon cloud of the polaron.

From the self-consistent sample for $G_{ii}(z)$ we can directly
read off the distribution for $\rho_i(\omega)$ in the form of
a histogram.
Instead of dealing with the full distribution
we can characterise its deviation from a Gaussian reasonably well 
by a `typical' value which is the geometric average of the random
quantity, i.e. 
\begin{equation}
\rho_{\rm ty}(\omega) = \exp \left[ \int \ln(\rho_i)
P[\rho_i(\omega)] \, d\rho_i \right]\,.
\end{equation}
In the following  
we classify therefore states at energy $\omega$ as localised (extended) 
if the typical value $\rho_{\rm ty}(\omega)$ 
vanishes (is finite) and 
the average $\rho_{\rm av}(\omega)$ is finite. The latter condition ensures that
there is a state at energy $\omega$.  

\section{Limiting cases}

For the free electron, with $\epsilon_i=0$ and $\Sigma_i=0$,
Eq.~(\ref{Gii}) is a quadratic equation whose solution provides the 
semi-circular DOS $\rho(\omega) = 
(4/\pi W_0^2) \sqrt{W_0^2- 4 \omega^2}$ of the Bethe lattice, with bare bandwidth 
$W_0=4 t \sqrt{K}$. 
In what follows, we set $W_0=1$ on a $K=2$ Bethe lattice and measure energies in units
of $W_0$, defining the dimensionless parameters
disorder strength $\bar{\gamma}=\gamma/W_0$,
EP coupling constants $\bar{\lambda}=2 \varepsilon_p/W_0$ and $g^2=\varepsilon_p/\omega_0$, 
and phonon frequency $\bar{\omega}_0= \omega_0/W_0$.

\subsection{Anderson model}

\begin{figure}[t]
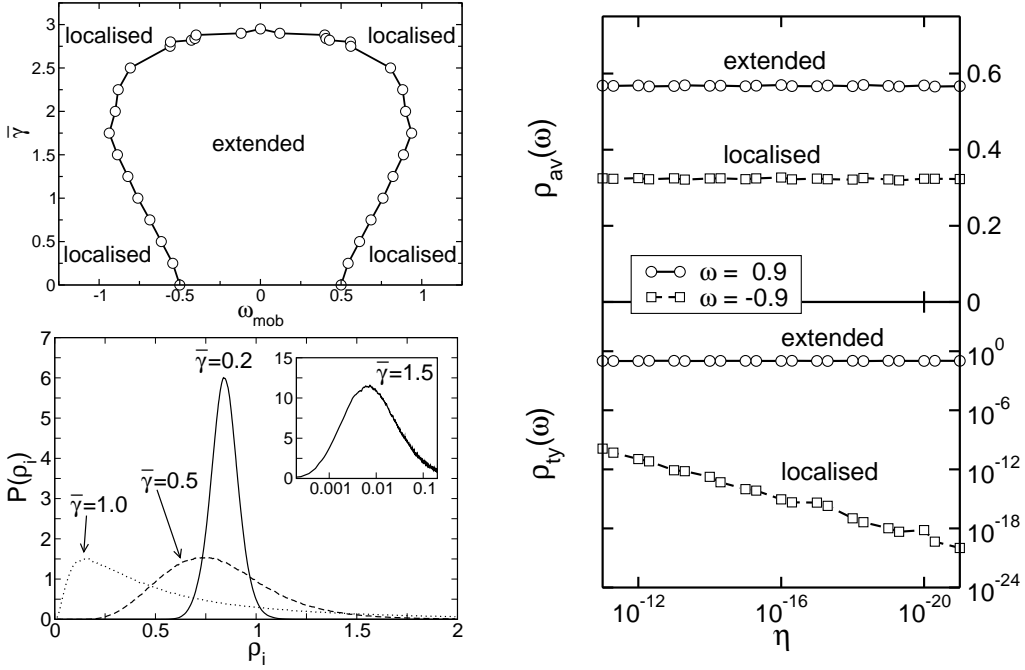

\begin{minipage}{0.45\linewidth}
\includegraphics[width=\linewidth]{fehske_ba_f3a} \\
\includegraphics[width=\linewidth]{fehske_ba_f3b} 
\end{minipage}\hfill
\begin{minipage}{0.49\linewidth}
\includegraphics[width=\linewidth]{fehske_ba_f3c}
\end{minipage}
\caption{\small The upper left panel depicts the mobility edge
trajectory for a single electron in the bare Anderson model.
The lower left panel gives the distribution $P[\rho_i(\omega=0)]$
for different $\bar{\gamma}$.
 The right panel shows the behaviour of the averaged and typical LDOS
 for extended and localised states in the limit $\eta\to 0$ ($\bar{\gamma}=1.5$).} 
\label{MEelectron}
\end{figure}

Let us first discuss the bare Anderson model ($\bar{\lambda}=0$), 
 when our approach reduces to the self-consistent theory
of localisation~\cite{AAT73}.
Then we can determine the
mobility edge trajectory for a single electron on the $K=2$ Bethe
lattice. 

For a qualitative discussion
of the localisation behaviour, it is sufficient to obtain random samples for a 
small but finite imaginary part of $z = \omega + \mathrm{i} \eta$.
If not stated otherwise, $\eta=10^{-8}$. 
For a quantitative determination of mobility edges it is necessary,
however, 
to track the average and typical LDOS for $\eta\to0$. The
fingerprint of a localised state at energy $\omega$ is a finite
$\rho_{\rm av}(\omega)$ and a vanishing $\rho_{\rm ty}(\omega)$ for
$\eta\to 0$ (see right panel in Fig.~\ref{MEelectron}). 
In the upper left panel of Fig.~\ref{MEelectron} we present the mobility
 edge trajectory for a single electron in the Anderson model obtained
 this way. 

Obviously, our approach is accurate enough to detect the two competing effects
associated with the characteristic re-entrance behaviour of the mobility edge
trajectory: The formation of delocalised states in the vicinity of the
band edge at small disorder and the localisation transition at large
disorder~\cite{KK93}.
 Even in the strongly disordered regime, where the mobility
edges move to the centre of the band and all states are localised, the
statDMFT works  reliably well. We did not attempt to estimate the
statistical error,  but it should be of the order of the fluctuations
visible in the plot. 
By its construction on a Bethe lattice, and in view of the localisation
behaviour obtained, the statDMFT comprehensively describes
localisation in a 3D system.
The localisation behaviour in 1D and 2D is, in accordance
with the goal of our studies, beyond the scope of this approach.

\subsection{Holstein model}

\begin{figure}[t]
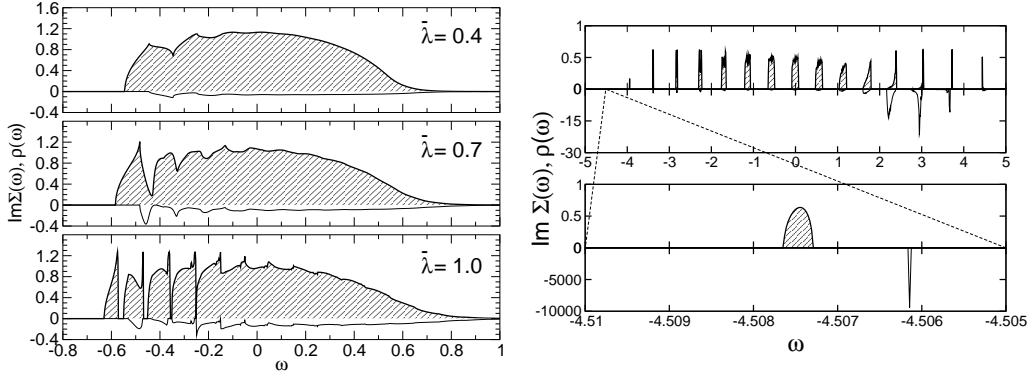

\begin{minipage}{0.49\linewidth}
\includegraphics[width=\linewidth]{fehske_ba_f4a.eps}
\end{minipage}\hspace{0.02\linewidth}
\begin{minipage}{0.49\linewidth}
\includegraphics[width=\linewidth]{fehske_ba_f4b.eps}
\end{minipage}
\caption{\small Left panel: Evolution of the DMFT DOS (thick solid line) and the 
imaginary part of the interaction self-energy (thin solid line) for the Holstein model 
with $\bar{\omega}_0=0.1$ and increasing EP coupling $\bar{\lambda}=0.4, 0.7$, and $1.0$.
Right panel: DMFT DOS and imaginary part of the interaction self-energy 
in the anti-adiabatic strong-coupling regime
($\bar{\omega}_0=0.5625$ and $\bar{\lambda}=9.0$). The lower panel shows the 
extremely narrow lowest sub-band ($W=3.45\times 10^{-4}$) in more detail.}
\label{Evol}
\end{figure}

For $\gamma=0$, our approach reduces to the DMFT of a single polaron in the Holstein 
model~\cite{Su74CDF97}, the physical properties of which are governed by 
two parameter ratios: The adiabaticity ratio $\bar{\omega}_0$
and the dimensionless EP coupling constants $\bar{\lambda}$ and $g^2$.
 Polaron formation sets in if both $\bar{\lambda}>1$ and $g^2>1$. 

The left panel of Fig. \ref{Evol} illustrates the on-set of polaron formation with increasing
EP coupling. The DOS fragments into an increasing number of sub-bands,
and a new energy scale arises: The strongly renormalised width $W$ of
the polaronic sub-bands.
As far as localisation is concerned, the lowest sub-band is particularly important, 
because it is completely coherent (${\rm Im}\,\Sigma_i(\omega)=0$) and inelastic 
scattering does not mask the localisation process.  
The internal structure of 
the states of the lowest sub-band depends on $\bar{\lambda}$ and
$\bar{\omega}_0$. For the adiabatic polaron, shown in the left panel of
Fig. \ref{Evol}, this gives rise to a  pronounced asymmetry in the DOS
(provided $\bar{\lambda}$ is not too large).
Specifically the states at the top of the lowest sub-band are phonon dominated 
leading to a substantial band flattening.
States at the bottom of the sub-band have a smaller phonon admixture. 
These states are more mobile as compared to a rescaled tight-binding electron
due to long-range tunnelling processes induced by EP coupling~\cite{WF98a}.
Both, band flattening and long-range tunnelling
will of course affect the localisation behaviour of a polaron.
In contrast, the composition of the states of the lowest sub-band is
more or less homogeneous in the anti-adiabatic strong-coupling regime 
shown in the right panel of Fig.~\ref{Evol} . Self-trapping leads here to a band collapse resulting in
an extremely narrow sub-band whose DOS is essentially identical to the
rescaled semi-circular DOS of the Bethe lattice.

\subsection{Infinite dimension}

DMFT becomes exact if the lattice has infinite connectivity.
In the limit $K\to\infty$ we have to re-scale the transfer amplitude 
as $t\rightarrow \bar{t}/\sqrt{K}$ to keep the
bandwidth $W_0=4 t \sqrt{K}$ constant.
Applying the central limit theorem to $H_i(z)$ we find 
that $H_i(z)$ converges to its arithmetic mean (average value).
 The configuration 
average of Eq.~(\ref{Gii}) is then only a simple single-site average over
$\epsilon_i$ and leads to a closed system of equations -- precisely the DMFT equations --
for the averaged local Green's function $G_\mathrm{av}(z)$. 
Each 
impurity site experiences in this limit the same environment.
Disorder is thus only 
treated at the level of the coherent potential approximation and
localisation cannot be detected.
Note that the statDMFT, instead, keeps $K$ finite. 
Hence, the environment of each impurity site is allowed to fluctuate leading
to Anderson localisation provided the fluctuations are strong enough. 

\section{Anderson-Holstein model}

We now turn our attention to the combined effect of disorder and EP 
coupling.
At weak enough EP coupling, where the mean free path due 
to inelastic EP scattering is shorter than the localisation length, 
EP coupling suppresses localisation~\cite{LR85}.
 Phonons act then only as scattering 
centres which randomly change the phase of the electron wave-function before 
destructive interference due to elastic electron-impurity scattering --
being essential for Anderson localisation -- can occur.
In the strong EP coupling regime, however, 
 the DOS of the clean system
fragments into polaronic sub-bands (Fig.~\ref{Evol}), the lowest of 
which turns out to be completely coherent. States comprising this sub-band 
suffer no inelastic scattering -- an ideal situation for destructive interference and
thus for Anderson localisation.
Because it is also much narrower than the bare electron band, extremely small amounts 
of disorder would suffice to localise states within this sub-band.

 Obviously, the spread $\bar{\gamma}$ of the on-site energies $\epsilon_i$ can be small on the
scale of the bare bandwidth but large on the scale of the width of the sub-bands. In the 
Anderson-Holstein model, we have to distinguish therefore at least between two regimes:
The {\it Holstein regime}, where disorder is small on the scale of the bare bandwidth,
and the {\it Anderson regime}, where it is large on the scale of the
width of the polaronic sub-bands.

\subsection{Anderson regime}
First we consider the Anderson regime.
In addition to the LDOS $\rho_i(\omega)$, the statDMFT suggests to use the 
imaginary part of the hybridisation function, the 
escape rate from a given site 
$\Gamma_{i}(\omega)= - \mathrm{Im}\, H_i(\omega) / \pi$,
to characterise the localisation behaviour of a state at energy 
$\omega$.  Obviously, a finite (vanishing) escape rate $\Gamma_i(\omega)$
implies an extended (localised) state at energy $\omega$. 
Similar to the LDOS $\rho_i(\omega)$ we have to consider its typical
value $\Gamma_{\rm ty}(\omega)$.

\begin{figure}[t]
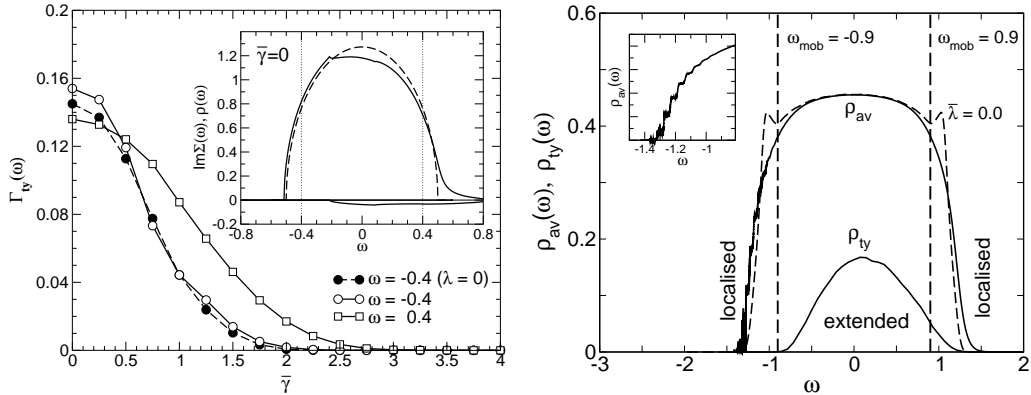

\begin{minipage}{0.49\linewidth}
\includegraphics[width=\linewidth]{fehske_ba_f5a.eps}
\end{minipage}\hspace{0.02\linewidth}
\begin{minipage}{0.49\linewidth}
\includegraphics[width=\linewidth]{fehske_ba_f5b.eps}
\end{minipage}
\caption{\small Left panel: Typical escape rate in the Anderson
regime for weak EP coupling ($\bar{\lambda}=0.067$ and $\bar{\omega}_0=0.3$). 
Filled circles denote data for $\bar{\lambda}=0$. The inset shows the
DMFT DOS and the imaginary part of the interaction self-energy for
$\bar{\gamma}=0$, as well as the
bare DOS for $\bar{\lambda}=\bar{\gamma}=0$ (dashed line). Vertical dotted lines
indicate the energies for which the escape rate is plotted.
Right panel: Typical LDOS and DMFT DOS in the Anderson regime 
for intermediate EP coupling ($\bar{\lambda}=1.0$, $\bar{\omega}_0=0.05$, and $\bar{\gamma}=2$).
The dashed line depicts the averaged LDOS for $\bar{\lambda}=0$ .
Vertical dashed lines indicate the mobility edges of the
noninteracting system;
the inset displays the DMFT DOS at the left band edge.}
\label{ARweak}
\end{figure}

The left panel of Fig.~\ref{ARweak} displays typical escape rates as 
a function of $\bar{\gamma}$ for two energies, one below and one above
the phonon emission threshold (see inset). 
Below the threshold, ${\rm Im}\Sigma_i(\omega)=0$, and the states are coherent. 
Anderson localisation occurs at the value
$(\bar{\gamma}_c)_{\omega=-0.4}\approx 2.25$ for which 
$\Gamma_{\rm ty}(\omega)$ vanishes.
Since the phonon dressing of these
states is negligible the escape rates of the interacting and 
non-interacting system vanish at roughly the same $\bar{\gamma}_c$ .
Above the threshold, 
inelastic EP scattering (${\rm Im}\,\Sigma_i(\omega)\neq 0$)
disrupts the destructive interference  required for 
localisation.
Here, the critical disorder strength
$(\bar{\gamma}_c)_{\omega=0.4}\approx 3.0$
for which $\Gamma_{\rm ty}(\omega)$ vanishes is larger
than below the phonon emission threshold.
Thus, above the threshold, localisation is 
suppressed. It is rather encouraging that our approach recovers this
basic feature of the physics of localisation.

In the right panel of Fig.~\ref{ARweak} we show the typical LDOS  
together with the DMFT DOS for $\bar{\omega}_0=0.05$, $\bar{\lambda}=1.0$,
and $\bar{\gamma}=2$. Without EP coupling, there would be mobility
edges at $\omega\approx\pm 0.9$ (see Fig. \ref{MEelectron}).
 In the presence of EP coupling, the lower mobility edge
is still located at the
mobility edge of the noninteracting system,
but the upper mobility edge is shifted to higher
energies.
Thus, precisely at the upper mobility edge EP coupling delocalises states.

To explain this asymmetry, recall that our calculation is for
$T=0$. Hence, states at energy $\omega$ can, due to EP interaction, only
couple to states at energies less than $\omega$. 
At the high energy side localised states above the
mobility edge of the noninteracting system couple to delocalised
states below. As a consequence, formerly localised states become delocalised. 
In contrast, states below the lower mobility edge of the noninteracting 
system remain localised, because they can only couple to states which
are already localised. 
Additionally, EP interaction
attempts to transform electronic band states above the lower
mobility edge of the noninteracting system into polaronic states,
as suggested by Anderson~\cite{An72}.
Hence, these states become heavier and
more susceptible to disorder. 
As a result,
at the lower
mobility edge EP coupling enhances the tendency towards
localisation.

Let us take a closer look at the DMFT DOS,
which at the low-energy side shows pronounced plateaus 
with a width given by the phonon energy. 
This step-like increase of the DOS, together with the vanishing of the 
typical LDOS, is a clear signature for localised polaronic defect
states.
Being strongly localised these states can be
described by the independent boson model whose density of states
consists of discrete peaks separated by $\omega_0$.
The DOS does not change with energy as long as the fluctuations
of the on-site energies are smaller than the phonon energy. If the difference
in on-site energy is equal to the phonon energy, a step arises because states
with one additional phonon contribute. The step height reflects therefore 
the phonon distribution of the polaronic defect states.

Our approach identifies the mobility edge  
which separates, in the strongly disordered Anderson regime, polaronic
defect states from itinerant polarons. This is a fundamental first step 
towards a microscopic theory of polaron transport in disordered media
which, in principle, could cover thermally activated hopping
between polaronic defects below the mobility edge.

\subsection{Holstein regime}

In the Holstein regime the spread of the on-site energies is much smaller than 
the bare bandwidth. We only consider the situation where $\bar{\gamma}$ is 
so small that disorder cannot mix different polaronic sub-bands. 
In that case, mobility edges within the 
lowest sub-band can be identified and tracked as a function of disorder. 

\begin{figure}[t]
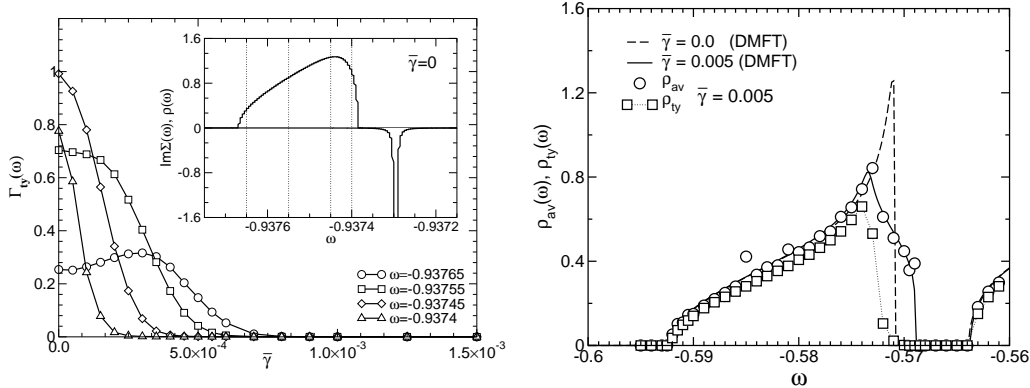

\begin{minipage}{0.49\linewidth}
\includegraphics[width=\linewidth]{fehske_ba_f6a}
\end{minipage}\hspace{0.02\linewidth}
\begin{minipage}{0.49\linewidth}
\includegraphics[width=\linewidth]{fehske_ba_f6b}
\end{minipage}
\caption{\small Left panel: Typical escape rates in the
adiabatic intermediate-to-strong coupling regime ($\bar{\omega}_0=0.1$,
$\bar{\lambda}=1.8$). The inset displays the DMFT
DOS for the lowest sub-band and the imaginary part of the interaction self-energy
for $\bar{\gamma}=0$; vertical dotted lines indicate the energies
for which $\Gamma_{\rm ty}(\omega)$ is plotted.
Right panel: Average and typical LDOS in the  polaron cross-over
regime ($\bar{\omega}_0=0.0625$, $\bar{\lambda}=1.0$) for $\bar{\gamma}=0.005$. Note,
the average LDOS is almost perfectly approximated by the DMFT DOS. At
about $\omega=-0.565$ the second polaronic sub-band starts.}
\label{HRescape}
\end{figure}

In the left panel of Fig.~\ref{HRescape} we present typical
escape rates as a function of $\bar{\gamma}$ for the lowest polaronic
sub-band.
In the adiabatic intermediate-to-strong coupling regime, the DMFT DOS (for $\bar{\gamma}=0$, inset of
Fig.~\ref{HRescape}) is rather asymmetric because of the
`hybridisation' with the (optical) phonon branch, leading to band
flattening. As a result, states at the
top of the sub-band are very susceptible to disorder and the critical
disorder strength for which $\Gamma_{\rm ty}(\omega)$ 
vanishes is substantially smaller than
at the bottom of the sub-band. Clearly, in contrast to the bare Anderson
model, there is a significant asymmetry of the localisation behaviour
of the states within the sub-band. 
To demonstrate the effect of this asymmetry more clearly, we plot in the right panel
of Fig. \ref{HRescape} the typical and average LDOS for the lowest
sub-band in the adiabatic cross-over regime for $\bar{\gamma}=0.005$. 
The steeple-like shape due to band flattening is now very pronounced and 
the suppression of the typical LDOS at the top of the sub-band 
can be clearly seen. 

\begin{figure}[t]
\begin{minipage}{0.4\linewidth}
\includegraphics[width=\linewidth]{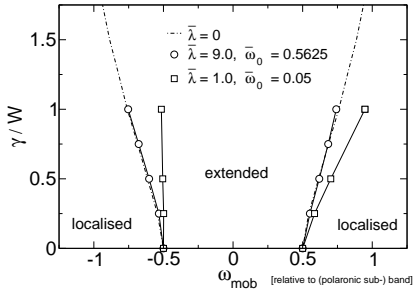}
\end{minipage} \hfill
\begin{minipage}{0.55\linewidth}
\caption{\small Parts of the mobility edge trajectory for the lowest polaronic
sub-band of the Anderson-Holstein model in the adiabatic cross-over 
($\bar{\omega}_0=0.05$ and $\bar{\lambda}=1$, $\square$) and the 
anti-adiabatic strong coupling regime ($\bar{\omega}_0=0.5625$ and 
$\bar{\lambda}=9$, $\circ$). Disorder is measured in units of the respective
width $W$ of the sub-bands, and $\omega_\mathrm{mob}$ is given relative to
the sub-band position.
The mobility edge trajectory for the bare Anderson model (dashed-dotted line) is given for comparison.}
\label{MEahm}
\end{minipage}
\end{figure}

The same asymmetry is manifest in 
the mobility edge trajectory
for the lowest sub-band in the adiabatic cross-over regime, 
which we partly show in Fig.~\ref{MEahm}.
As expected, states at the bottom of the
sub-band are almost insensitive to small amounts of disorder.
The lower mobility edge (and also the lower band edge) is essentially
pinned.
Notice, in this regime 
polaron states at the bottom of the
sub-band are even more difficult to localise than bare electron states
if we measure the disorder on the relevant energy scale $W$ or $W_0$,
respectively.
Thus, even in absence of the suppression of localisation due to inelastic
scattering, a polaron is not always easier to localise than an electron. 
States at the top of the sub-band, on the other hand, are immediately 
affected by disorder. Small amounts of disorder (small even on the scale of
the sub-band) are sufficient to localise the states,
and the upper mobility edge rapidly shifts to higher energies. 

For (large) disorder disorder $\gamma\gtrsim W$, 
the upper band edge of the disorder-broadened
lowest sub-band moves into a spectral range with significant
inelastic scattering (not shown in Fig. \ref{MEahm}). 
The enhanced inelastic scattering rate for the
states at the top of the lowest sub-band strongly
suppresses localisation in this spectral range.
With increasing disorder states move into the gap region,
which separates the first from the second sub-band, thereby giving
rise to a shrinking gap. Finally both sub-bands merge. 
This redistribution of states occurs before a re-entrance behaviour of the
mobility edge trajectories can be observed. 

Also shown in Fig.~\ref{MEahm} is the mobility edge trajectory
for the lowest sub-band in the anti-adiabatic strong-coupling regime
($\bar{\omega}_0=0.5625$ and $\bar{\lambda}=9$). 
Here self-trapping manifests itself in a very narrow coherent
band which is essentially a rescaled replica of the bare band.
The localisation properties of the lowest sub-band in the anti-adiabatic
strong-coupling regime are therefore those of a rescaled Anderson
model as can be seen from the perfect match of the properly scaled
mobility edge trajectories. 

\section{Conclusions}

Based on a stochastic Green's function approach we analysed, 
for representative situations, the interplay between Anderson 
localisation and self-trapping in the framework of the Anderson-Holstein model.

Only in the anti-adiabatic Holstein regime the localisation properties
are essentially those of a rescaled Anderson model whereas they are
strongly affected by the internal structure of the phonon dressing in
all other cases. 
As demonstrated EP coupling can either work against localisation due
to inelastic scattering, or enhance the tendency towards localisation
due to self-trapping. 
A polaron can be rather easy to localise with respect to the
energy scale set by EP coupling (e.g. the width of the lowest
polaronic sub-band) -- but notably, also in regions without inelastic
scattering, it can be even harder to localise than the bare electron. 
Most importantly, EP coupling can change localised electron states 
to strongly localised polaronic defect states. Transport 
between these states has to occur by thermally activated hopping. Our 
approach, which is capable to track this transmutation, 
should be thus suitable for investigating the cross-over from 
hopping transport below the lower mobility edge to band transport
above~it.

Our investigation of the Anderson-Holstein model is far from being complete. 
For instance, a systematic study of 
the parameter range where disorder mixes polaronic sub-bands has not yet been
done, as well as an investigation at finite temperatures or  
particle densities. In particular the latter is rather complicated because the 
EP self-energy can then be no longer analytically obtained in the form of
a continued fraction. Most pressing, however, is the calculation of the 
finite temperature conductivity for a single polaron 
because it would open the door to a microscopic theory of polaron transport.
Preliminary work in this direction exists only without disorder~\cite{FC03}, 
which is however crucial for a complete description of 
polaron transport.



\end{document}